\documentclass[10pt]{amsart}
\usepackage{amssymb,amsthm,euscript,amsmath,bm,graphicx}
\usepackage{times}
\usepackage{amsfonts}

\newcommand{\be}{\begin{equation}}
\newcommand{\ee}{\end{equation}}
\newcommand{\ben}{\begin{equation*}}
\newcommand{\een}{\end{equation*}}
\newcommand{\bea}{\begin{eqnarray}}
\newcommand{\eean}{\end{eqnarray*}}
\newcommand{\bean}{\begin{eqnarray*}}
\newcommand{\eea}{\end{eqnarray}}

\newcommand{\ba}{\boldsymbol\alpha}

\newcommand\re[1]{(\ref{#1})}

\begin{document}

\title{Internal variables and dynamic degrees of freedom}
\author{\bf P. V\'an*, A. Berezovski and J. Engelbrecht}

\address{Centre for Nonlinear Studies\\Institute of Cybernetics at Tallinn University of Technology\\
12618 Tallinn, Akadeemia tee 21, Estonia, \\
*Department of Theoretical Physics, Research Institute for Particle
  and Nuclear Physics, HAS,\\
H-1525, Budapest, Konkoly Thege M. \'ut 29-33, Hungary\\ and\\
Bergen Computational Physics Laboratory, BCCS,\\
N-5008 Bergen, Thormohlensgate 55 Norway. }

\email{vpet@rmki.kfki.hu}

\date{\today}


\begin{abstract} Dynamic degrees of freedom and internal variables are treated in a uniform way.
The unification is achieved by means of the introduction of a dual
internal variable. This duality provides the corresponding evolution
equations depending on whether the Onsager-Casimir reciprocity
relations are satisfied or not.
\end{abstract}

\maketitle

Running title: Internal variables and dynamic degrees of freedom

\newpage

\section{Introduction}

Introduction of internal variables suggests taking into account the
influence of an internal structure on the dynamic behaviour of a
material \cite{Ric71a}. As long as evolution equations of internal
variables are determined by general macroscopic principles, we can
expect that the validity of the evolution equations is independent
of particular microscopic models. It is important, therefore, to
understand the universal aspects behind these evolution equations.

There are two basic methods to generate the evolution equations for
internal variables. Both methods are based on fundamental
principles.

The {\em first method} generates the evolution equations exploiting
the entropy inequality. This approach uses exclusively thermodynamic
principles, and the corresponding variables are called {\em internal
variables of state} \cite{Mau06a}. This frame has the advantage of
operating with familiar thermodynamic concepts (thermodynamic force,
entropy), however, no inertial effects are considered. The
thermodynamic theory of internal variables has a rich history (see
the historical notes in \cite{MauMus94a1}). A first more or less
complete thermodynamic theory was suggested by Coleman and Gurtin
\cite{ColGur67a}, and the clear presentation of the general ideas of
the theory was given by Muschik \cite{Mus90a1}. Internal variables
were applied for several phenomena in different areas of physics,
biology, and material sciences. A complete description of the
thermodynamic theory with plenty of applications based on this
concept of internal variables of state can be found in
\cite{Ver97b}.

The {\em second method} generates the kinetic relations through the
Hamiltonian variational principle and suggests that inertial effects
are unavoidable. This approach has a mechanical flavor, and the
corresponding variables are called {\em internal degrees of
freedom}. Dissipation is added by dissipation potentials. This
theoretical frame has the advantage of operating with familiar
mechanical concepts (force, energy). The method was suggested by
Maugin \cite{Mau90a}, and it also has a large number of applications
\cite{Mau99b,Eng97b}. The clear distinction between these two
methods with a number of application areas is given by Maugin and
Muschik \cite{MauMus94a1,MauMus94a2} and Maugin \cite{Mau06a}.

It is important to remark that the terminology of the field is not
unique, and even more, it is contradictory. For example, Verh\'as
\cite{Ver97b} calls his internal variables as dynamic degrees of
freedom following the suggestion of Kluitenberg (see e.g.
\cite{Klu77a,KluCia78a}). Sometimes internal variables or internal
degrees of freedom appear without thermodynamic foundation under
different names. An important example can be found in nerve pulse
dynamics, where the classical "phenomenological variables" of
Hodgkin and Huxley \cite{HodHux52a} and the "recovery variables" of
Nagumo, Arimoto and Yoshizawa \cite{NagAta62a,Izh00a} are internal
variables from a thermodynamic point of view, as it was shown in
\cite{MauEng94a}. Another important example is damage mechanics
(e.g. \cite{Kra96b}), where the macroscopic damage variable is an
internal variable from a thermodynamic point of view, and the
thermodynamic framework can reveal several new properties of the
theory \cite{MurKam96a,Van01a1}.

In this paper we follow the terminology of Maugin and Muschik
\cite{MauMus94a1} with some important extensions. We call internal
variables of state those physical field quantities - beyond the
classical ones - whose evolution is determined by thermodynamical
principles. We call internal degrees of freedom those physical
quantities - beyond the classical ones - whose dynamics is
determined by mechanical principles.

One of the questions concerning this doubled theoretical frame is
related to common application of variational principles and
thermodynamics. Basic physical equations of thermodynamical origin
do not have variational formulations, at least without any further
ado \cite{VanMus95a}. That is well reflected by the appearance of
dissipation potentials as separate theoretical entities in
variational models dealing with dissipation.

On the other hand, with pure thermodynamical methods - in the
internal variables approach - inertial effects are not considered.
Therefore, the coupling to simplest mechanical processes seemingly
requires to introduce some improvements, those are usually new
principles of mechanical origin.

Therefore, one can have the impression that the doubling of the
theoretical structure is a necessity, because the usage of both
mechanical and thermodynamical principles cannot be avoided. This is
the conceptual standpoint of the GENERIC approach (GEneral
Non-equilibrium Equation of Reversible Irreversible Coupling)
\cite{Ott05b}. However, a doubled theoretical structure is not very
convenient, because it doubles the number of physical assumptions
restricting predictive capabilities of the theory.

In what follows, we propose a uniform approach  based exclusively on
thermodynamic principles. Our suggestion requires dual internal
variables and a generalization of the usual postulates of
non-equilibrium thermodynamics: we do not require the satisfaction
of the Onsagerian reciprocity relations. With dual internal
variables we are able to include inertial effects and to reproduce
the evolution of dynamic degrees of freedom. It could be impossible
with a single internal variable. This is the price we pay for the
generalization. In other words, instead of the doubling of the
theoretical structure we suggest the doubling of the number of
internal variables.

In the next Section we give a short overview of the thermodynamic
frame, then we summarize the theory of internal variables and
internal degrees of freedom in Sections 3 and 4, respectively. The
unified frame is developed in Section 5. Finally, we shortly discuss
the advantages and disadvantages of the suggested approach.

\section{Thermodynamics of continua}

First, we recall the basic equations of the finite-strain
thermoviscoelastic continua with internal variables. In this case,
the state space of a material point $\bf X$ is spanned by the
variables $(e,{\bf F},\ba)$, where $e$ is the specific internal
energy, {\bf F} is the deformation gradient (direct-motion gradient)
defined as ${\bf F} = \frac{\partial {\bf x}}{\partial {\bf
X}}\left.\right|_t$, and $\ba= (\alpha_1, \alpha_2, ... ,\alpha_n)$
are internal variables of state. Here ${\bf x}= {\bf x}({\bf X},t)$
denotes the mapping from the reference configuration to an actual
one.

The balance of linear momentum in an actual configuration reads
 (in the absence of body forces) \cite{TruNol65b}
 \be
 \rho \dot{\bf v} - \nabla\cdot {\bf t} = {\bf 0},
\ee where ${\bf v}=\frac{\partial {\bf x}}{\partial
t}\left.\right|_{\bf X}$ is the velocity field, the dot above the
velocity denotes the substantial time derivative, and ${\bf t}$ is
the Cauchy stress. The density $\rho$ is not independent of the
deformation gradient, because the density in the reference
configuration is $\rho_0 = \rho \det {\bf F}=const$.

The balance of internal energy can be calculated as the difference
between the conserved total energy and the kinetic energy and is
given as (e.g. \cite{Ver97b})
 \be
 \rho \dot{e} + \nabla\cdot{\bf q} = {\bf t}:\dot{\bf F}{\bf F}^{-1}.
 \label{balen}\ee
Here ${\bf q}$ is the heat flux and $\dot{\bf F}{\bf F}^{-1} =
\nabla {\bf v}$. Then the entropy balance can be represented in the
form \cite{Ver97b,TruNol65b}
 \be
 \rho\dot{s} + \nabla\cdot {\bf j} = {\bf q}\cdot\nabla\frac{1}{T} +
 \frac{1}{T}\left({\bf t}- {\bf t}^{rev}\right):\dot{\bf F}{\bf
 F}^{-1} - \frac{1}{T}{\bf A}\cdot \dot\ba.
 \label{bale}\ee
Here we introduced the following traditional thermoelastic
quantities (given by Gibbs's equation):
 \be
 \frac{1}{T} = \frac{\partial s}{\partial e}, \quad
 {\bf t}^{rev} = \rho T {\bf F}\frac{\partial s}{\partial {\bf F}},
 \ee
  \be
 {\bf A} = (A_1, A_2, ..., A_n) =
    \left(-\rho\frac{\partial s}{\partial \alpha_1},
    -\rho\frac{\partial s}{\partial \alpha_2}, ...,
    -\rho\frac{\partial s}{\partial \alpha_n} \right).
 \label{aff}\ee
The entropy $s$ as a function of the state space variables $(e,{\bf
F}, \ba)$ is a concave function according to thermodynamic stability
requirements.

It is important to emphasize that the entropy flux {\bf j} in Eq.
\re{bale} is assumed to have its classical form
 \be
 {\bf j}_{cl} = \frac{\bf q}{T}.
 \ee
This is a severe assumption, and in the case of mixtures and weakly
nonlocal extension of classical theories the entropy flux deviates
from this classical form as $ {\bf j} = \frac{\bf q}{T} + {\bf J}$.
This additional term was proposed first by M\"uller \cite{Mul67a}.
The particular form of the additional term {\bf J} depends on the
thermodynamic interactions and their level of nonlocality (e.g., in
the case of weakly nonlocal heat conduction a family of its
different forms is presented in \cite{CimVan05a}).

In what follows we remind the derivation of the evolution equations
of internal variables from general principles. As the coupling to
mechanical or thermal interactions could be involved and would blur
the structure of equations, in the following we restrict ourselves
to the case, where the internal variables are completely decoupled
from the mechanical and thermal interactions. Namely, this is the
situation where the mechanical interaction is reversible ${\bf t} =
{\bf t}^{rev}$, the heat flux is zero ${\bf q}={\bf 0}$,
$\rho=const.$, ${\bf v} = const.$, and there is a complete
decoupling at the static level as well, that is $s(e,{\bf F}, \ba) =
s_T(e)+s_M({\bf F})+ s_A(\ba)$. Let us remark that one can give very
different conditions of decoupling, especially if other
thermodynamic potentials are introduced, as, e.g., the Helmholtz
free energy.

\section{Internal variables of state}

Let the basic thermodynamic state be spanned by $n$ scalar internal
variables $\ba=(\alpha_1,\alpha_2, ...,\alpha_n)$. We want to
determine their evolution according to thermodynamic principles.

\subsection{Local state}

Let us assume that the evolution equation of $\ba$ is given in a
general form
 \be
 \dot\ba =\boldsymbol{g}_\alpha,
\label{da}\ee where the dot denotes the material time derivative
that can be regarded as a partial one in our investigations.

The {\em local state hypothesis of Kestin} \cite{Kes93a1,Mus93a1} is
introduced by the assumption that entropy and the right hand side of
the evolution equation \re{da} are functions of $\ba$ and do not
depend on the derivatives of the basic state. Furthermore, we assume
that entropy has a maximum in thermodynamic equilibrium, that is
$\ba={\bf 0}$ at equilibrium. According to the Second Law, the
entropy function is negative semidefinite and increasing along the
processes determined by Eq. \re{da}. Therefore, the entropy balance
is given as
 \be
  \dot{\rho_s}+ \nabla\cdot {\bf J} = \sigma_s \geq 0.
 \label{eb1}\ee
Here $\rho_s = \rho s$ is the entropy density, ${\bf J}={\bf j} -
\frac{\bf q}{T}$ is the extra entropy flux and $\nabla\cdot$ is the
divergence operator.

Dissipation  can be calculated according to Eq. \re{bale}
 \be
 T\sigma_s =  -{\bf A}\boldsymbol{\cdot g}_\alpha =
    -\sum_{i=1}^n{A}_i g^i \geq 0,
\label{des}\ee
 where  the thermodynamic affinity ${\bf A}$
conjugated to the internal variable $\ba$ is determined by Eq.
\re{aff} and $ \boldsymbol{\cdot}$ denotes a duality mapping onto
the space of the internal variables as it is given by indexes.

\subsection{Fluxes and forces}

We may recognize a simple force-flux structure in Eq. \re{des} and
identify ${\bf A}$ as a thermodynamic force and ${\bf g}_\alpha$ as
a thermodynamic flux. A general solution of this inequality can be
given in the form of so called quasilinear \emph{conductivity
equations} \cite{Gur96a,Van03a}
 \be
 \boldsymbol{g}_\alpha(\ba) = -{\bf L}({\bf A}(\ba), \ba) {\bf A}(\ba).
\label{co}\ee Substituting the latter into Eq. \re{da} results in a
relaxation dynamics, because the conductivity matrix ${\bf L}$ is
positive semidefinite, as a consequence of Eq. \re{des}. We should
make a distinction of this quasilinear case, where the conductivity
matrix is a function of the forces and the state space (${\bf
L}({\bf A}(\ba), \ba)$), and of the strictly linear case, where the
conductivity matrix is constant (${\bf L} = const$). The strictly
linear theory is always an approximation that can be valid only in
the vicinity of equilibrium.

The negative sign of the affinity is traditional and can be
understood from the standard equilibrium formulation.

\subsection{Dissipation potentials}

General dissipation potentials were introduced by Onsager
\cite{Ons31a1,Ons31a2} in the case of strictly linear conductivity
equations. From the point of view of the general quasilinear
conductivity equations \re{co}, the existence of dissipation
potentials is connected to the validity of the general Gyarmati-Li
reciprocity relations \cite{Gya61a,Li62a}. In our case, the
Gyarmati-Li reciprocity relations in the so-called force
representation \cite{Gya70b} require that the derivative of the
thermodynamic fluxes ${\bf g}_\alpha$ with respect to the
thermodynamic forces (affinities ${\bf A}$), i.e.,
$\frac{\partial\boldsymbol{g}_\alpha}{\partial {\bf A}}({\bf
A},\ba)$ should be symmetric. This means that there exists a
dissipation potential $D_A({\bf A},\ba)$ with the property
$$
\frac{\partial D_A}{\partial{\bf A}}({\bf A},\ba) =
\boldsymbol{g}_\alpha({\bf A},\ba).
$$
Formally, assuming a symmetric relation of the thermodynamic fluxes
and forces, one can introduce the so-called flux representation
\cite{Gya70b} supposing that the thermodynamic forces are functions
of the thermodynamic fluxes. In this case, the condition of the
existence of the dissipation potential is connected to the validity
of the related Gyarmati-Li reciprocity relations, that is to the
symmetry of the derivative $\frac{\partial{\bf G_\alpha}}{\partial
A}(\boldsymbol{g}_\alpha,\ba)$. As a result, we can get a flux
related dissipation potential $D_g$ with the property
 $$
  \frac{\partial D_g}{\partial \boldsymbol{g}_\alpha}(\boldsymbol{g}_\alpha,\ba) =
  {\bf A}(\boldsymbol{g}_\alpha,\ba).
$$
In the strictly linear theory with a constant conductivity matrix
{\bf L}, these two representations can be transformed into each
other and both conditions coincide. Moreover, in this case the
general Gyarmati-Li reciprocity relations become equivalent to the
Onsagerian reciprocity relations.
In the quasilinear case, they are not equivalent. 
It is also important to remark that the role of fluxes and forces is
not completely interchangeable, because the fluxes are constitutive
functions, and the forces are given functions of the constitutive
state. Therefore, the validity of the flux representation is
restricted mostly to the strictly linear theory.

\subsection{Weakly non-local internal variables}

We now release the local state hypothesis and assume that the
entropy is a function of both the internal variables and their
gradients $(\ba,\nabla\ba)$. Moreover, one can postulate
(\cite{MauMus94a1}, \cite{Mau99b}) that the extra entropy flux is
 \be
 {\bf J} = -\left(\frac{\partial \rho_s}{\partial
 \nabla{\ba}}\right)\boldsymbol{g}_\alpha.
 \label{ec}\ee
In this case, the entropy balance \re{eb1} together with the
evolution equation \re{da} results in
 \bea
 \dot{\rho_s}(\ba,\nabla\ba) &+& \nabla\cdot{\bf J} = \nonumber\\
 &=&\frac{\partial \rho_s}{\partial\ba}\boldsymbol{\cdot}\dot{\ba} +
    \frac{\partial \rho_s}{\partial\nabla\ba}\nabla\dot{\ba} -
     \nabla\cdot \left(\frac{\partial \rho_s}{\partial\nabla\ba} \boldsymbol{g}_\alpha\right) = \nonumber\\
&=&  \left(\frac{\partial \rho_s}{\partial\ba} -
    \nabla\cdot\frac{\partial \rho_s}{\partial\nabla\ba}\right)\boldsymbol{\cdot g}_a  =
  -\verb"A"\boldsymbol{\cdot g}_\alpha \geq 0.
\label{epn}\eea Now the thermodynamic flux is the same as in the
local case, but the thermodynamic force has been changed to the new
weakly non-local affinity $\verb"A"= -\frac{\partial
\rho_s}{\partial\ba} + \nabla\cdot\frac{\partial \rho_s}{\partial
\nabla\ba}$. Therefore the evolution equation with quasilinear
conductivity is
 \be
 \dot{\ba} = {\bf L}\left(\frac{\partial \rho_s}{\partial\ba} -
    \nabla\cdot\frac{\partial \rho_s}{\partial \nabla\ba}\right).
 \label{GLev}\ee
It is remarkable that the right hand side of the evolution equation
become nonlocal as well.

The forces and the fluxes are compared in the following way:
 \vskip 3mm
\begin{center}\begin{tabular}{cccc}
Local force:  \hfill &
    $ \frac{\partial \rho_s}{\partial\ba}(\ba)$, \hfill
    & Nonlocal force: \hfill
    & $\frac{\partial \rho_s}{\partial\ba}(\ba,\nabla\ba) -
        \nabla\cdot\frac{\partial \rho_s}{\partial\nabla\ba}(\ba,\nabla\ba)$,\\
\hfill &\hfill\\
Local flux: \hfill &
    $\boldsymbol{g}_a(\ba)$, \hfill  &
    Nonlocal flux: \hfill&
    $\boldsymbol{g}_a(\ba,\nabla\ba,\nabla^2\ba).$
\end{tabular}\end{center}\vskip 3mm
Dissipation potentials can be generated as previously, but the
previous local affinity ${\bf A}$ should be changed to the new
weakly non-local affinity $\verb"A"$.

\section{Internal degrees of freedom}

Evolution equations of other microstructural variables - the {\em
internal degrees of freedom} - are generated by mechanical
principles. It is assumed that the dynamics of an internal degree of
freedom $\ba$ is determined by a variational principle of
Hamiltonian type with the Lagrangian
$\mathcal{L}(\dot\ba,\ba,\nabla\ba)$ and, therefore, it is governed
by a field equation of the canonical form (e.g. \cite{Mau99b}):
 \be
 \frac{\delta \mathcal{L}}{\delta \ba} =
  \frac{\partial \mathcal{L}}{\partial \ba} -
     \frac{d}{d t}
        \left(\frac{\partial \mathcal{L}}{\partial \dot{\ba}}\right) -
     \nabla\cdot\left(\frac{\partial \mathcal{L}}{\partial
        (\nabla \ba)}\right) = \bf{f}_\alpha.
\label{me}\ee Here ${\bf f}_\alpha$ is the dissipative force and
$\frac{\delta}{\delta\ba}$ is a functional derivative.

The Lagrangian $\mathcal{L}$ is usually divided into a kinetic and a
potential part as follows:
 \be
 \mathcal{L}(\dot{\ba}, \ba, \nabla\ba) = K(\dot{\ba}) - V(\ba, \nabla\ba).
\label{sl}\ee Substituting this relation into Eq. \re{me}, one can
get the following particular form of the evolution equation for the
internal degree of freedom $\ba$
 \be
\frac{d^2 K}{d\dot{\ba}^2} \ddot{\ba} =- \frac{\partial V}{\partial
\ba}
    + \nabla\cdot \frac{\partial V}{\partial \nabla\ba} -
    {\bf f}_\alpha,
 \label{Ne}\ee
where we have rearranged the terms to get an apparent Newtonian
form.

One can consider higher order space derivatives and generate
theories with higher order non-locality. On the other hand, one can
apply the local state hypothesis, where the Lagrangian does not
depend on the space derivatives and the above equation simplifies.
The natural boundary conditions for the internal degree of freedom
are consequences of the variational formulation:
 \be
 {\bf T}=\left.\frac{\partial \mathcal{L}}{\partial
    \nabla\ba}\right|_{\partial V} \cdot {\bf n},
 \label{bou}\ee
where  {\bf n} is the normal of the boundary of the considered
region $\partial V$ at a given point and ${\bf T}$ is the surface
"force" acting on the field $\ba$. The presence and the nature of
natural boundary conditions is sometimes connected to the
observability and controllability of the internal degree of freedom
\cite{MauMus94a1}.

Let us observe that the above mechanical description of the dynamics
of the internal degree of freedom  postulates a Hamiltonian
variational principle and immediately generates a second order
differential equation in time \re{Ne}. However, there is a natural
way to get two first order differential equations instead of a
second order one introducing the Hamiltonian through a suitable
Legendre transformation. This transformation reveals some internal
symmetries of the whole variational structure.

In fact, introducing a generalized momentum
$$
{\bf p}_\alpha = \frac{\partial \mathcal{L}}{\partial \dot{\ba}},
$$
and a Hamiltonian $H$, defined by the partial Legendre
transformation,
$$
{\bf p}_\alpha \dot\ba = \mathcal{L}(\dot\ba, \ba, \nabla\ba) +
    H({\bf p}_\alpha, \ba, \nabla\ba),
$$
we arrive at the first Hamiltonian equation
 \be
 \dot\ba = \frac{\partial H}{\partial {\bf
p}_\alpha}. \label{H1}
 \ee
On the other hand, the field equation \re{me} can be transformed
resulting in the second Hamiltonian equation
 \be
\dot{\bf p}_\alpha = - \frac{\partial H}{\partial \ba} +
     \nabla\cdot\left(\frac{\partial H}{\partial
        (\nabla \ba)}\right) - \bf{f}_\alpha.
 \label{H2}\ee
Let us consider a particular form of the Hamiltonian equations in
the case of the special Lagrangian given in Eq. \re{sl}. For
simplicity, we also assume  that the kinetic term $K$ is quadratic,
i.e. $K(\dot \ba) = m\dot{\ba}^2/2$, where $m=const$. In this case,
the generalized momentum is ${\bf p}_\alpha = m\dot {\ba}$, and one
can get $\dot {\ba}({\bf p}_\alpha)= {\bf p}_\alpha /m$. The
Hamiltonian follows as $H(\alpha,{\bf p}_\alpha) = \frac{{\bf
p}_\alpha^2}{2 m} + V({\ba},\nabla{\ba})$. Finally, the
corresponding special Hamilton equations are
 \bea
  \dot \ba &=& \frac{{\bf p}_{\ba}}{m}, \label{sh1}\\
  \dot{\bf p}_{\ba} &=& -\frac{\partial V}{\partial \ba} +
     \nabla\cdot\left(\frac{\partial V}{\partial
        (\nabla \ba)}\right) - \bf{f}_\alpha.
\label{sh2} \eea

\subsection{Dissipation potentials}

Some thermodynamic background can be added through the observation
that the right hand side of Eq. \re{me}, the so-called {\em
dissipative force} $\bf{f}_\alpha$, enters the entropy production
because it generates dissipated power. The external mechanical power
is included into the balance of total energy, then in the balance of
internal energy \re{balen}, and, as a consequence, in the entropy
balance \re{bale}, as well (an other reasoning is based on the
principle of virtual power \cite{Mau99b,Mau80a}). Finally, the
entropy production can be written as
 $$
T\sigma_s = {\bf f}_\alpha \boldsymbol{\cdot}\dot{\ba} \geq 0.
$$
The whole thermodynamic formalism of the previous section can be
repeated on the basis of the latter form of the entropy production,
but now the thermodynamic fluxes are dissipative forces. However,
here the thermodynamic formalism is added to the previous
considerations of mechanical origin. For example, it is convenient
to introduce dissipation potentials (with the above mentioned
conditions) $D_f$ in the flux representation and put the dynamic
equation \re{me} into the following form \cite{MauMus94a1}
 $$
 \frac{\delta \mathcal{L}}{\delta \ba} =
  \frac{\partial D_f}{\partial {\bf f}_\alpha}.
 $$

\section{Dual internal variables}

As we have seen, evolution equations for internal variables of state
and for internal degrees of freedom are completely different (cf.
Eqs. (\ref{da}) and (\ref{Ne})). The same is clearly demonstrated
recently on the example of continuum thermomechanics \cite{Mau06a}.

The question arises whether it is possible to construct a single way
of the derivation of evolution equations for both internal variables
of state and internal degrees of freedom, or not?

A first guess may be that the internal variables of state are
special cases of the internal degrees of freedom, because it looks
like very easy to get a first-order time-derivative equation from
the second-order one. However, this is not the case. On the
contrary, the structure of Hamiltonian differential equations
\re{H1} and \re{H2} is very special. The essential part of the
problem is that a second-order time-derivative equation can be
generated by a Hamiltonian variational principle similar to the
traditional variational principle in mechanics without any further
ado, but a first-order time-derivative equation only in special
constrained cases (gyroscopic degeneracy \cite{MarRat99b}). This is
a strong mathematical restriction, and any attempt to circumvent the
problem has a price of loosing some parts of the nice Hamiltonian
structure \cite{VanMus95a,VanNyi99a}.

Another important observation is that the dissipative part of the
dynamics is generated by dissipation potentials also in the second,
mechanical method. It is remarkable that the mechanical generation
of evolution exploits thermodynamical methods.

In order to answer the formulated question,  we introduce a dual
internal variable and compare the arising evolution equations with
the corresponding equations obtained in the previous sections.

Let us consider a thermodynamic system where the state space is
spanned by two scalar internal variables $\alpha, \beta$. Then the
evolution of these variables is determined by the following
differential equations
 \bea
 \dot{\alpha}&=&g_\alpha, \label{d1}\\
 \dot{\beta}&=&g_\beta, \label{d2}
 \eea
where the functions $g_\alpha$ and $g_\beta$ of the right hand side
of the differential equations are constitutive functions and should
be restricted by the Second Law of thermodynamics. The entropy
inequality, the main ingredient of the Second Law, is the same as
previously (Eq. \re{eb1}). Let the domain of the constitutive
functions (our constitutive space) is spanned by the state space
variables and by their first and second gradients. Therefore, our
constitutive functions $g_\alpha, g_\beta, {\bf J}$ and $\rho_s$ are
given as functions of the variables $\alpha, \nabla\alpha,
\nabla^2\alpha, \beta, \nabla\beta, \nabla^2\beta$. This is a weakly
nonlocal constitutive space with second order weak non-locality in
both variables.

\subsection{Liu procedure}

The Second Law restricts the form of the possible evolution
equations (Eqs. \re{d1}-\re{d2}) and several exploitation methods of
the Second Law can be implemented \cite{MusEhr96a}. Here we apply
the procedure of Liu \cite{Liu72a}.

The gradients of the constitutive functions $g_\alpha$ and $g_\beta$
are constrained by the entropy inequality in the framework of a
second order constitutive state space for both of our variables as
follows \cite{Van05a}:
 \bea
 \nabla\dot{\alpha}&=&\nabla g_\alpha(\alpha, \nabla
\alpha, \nabla^2\alpha, \beta, \nabla\beta, \nabla^2\beta), \label{dd1}\\
 \nabla\dot{\beta}&=&\nabla g_\beta(\alpha, \nabla
\alpha, \nabla^2\alpha, \beta, \nabla\beta, \nabla^2\beta).
\label{dd2}
 \eea
Let us introduce the Lagrange-Farkas multipliers $\Lambda_\alpha,
\Lambda_{\nabla\alpha}, \Lambda_\beta, \Lambda_{\nabla\beta}$ for
Eqs. \re{d1}, \re{dd1}, \re{d2} and \re{dd2}, respectively.
Developing the partial derivatives of the constitutive functions, we
use the the numbering of the variables of the constitutive state
space spanned by $(\alpha, \nabla\alpha, \nabla^2\alpha, \beta,
\nabla\beta, \nabla^2\beta)$ as a convenient and short notation,
e.g., $\partial_3 \rho_s = \frac{\partial \rho_s}{\partial \nabla^2
\alpha}$. Therefore, Eq. \re{eb1} can be written as follows:
 \begin{gather}
 \dot{\rho_s} + \nabla\cdot{\bf J}
 - \Lambda_\alpha (\dot\alpha - g_\alpha)
 - \Lambda_{\nabla\alpha} (\nabla\dot\alpha - \nabla g_\alpha)
 - \Lambda_\beta (\dot\beta - g_\beta)
 - \Lambda_{\nabla\beta} (\nabla\dot\beta - \nabla g_\beta) = \nonumber\\
\partial_1\rho_s\dot\alpha +
    \partial_2\rho_s\nabla\dot\alpha +
    \partial_3\rho_s\nabla^2\dot\alpha +
    \partial_4\rho_s\dot\beta +
    \partial_5\rho_s\nabla\dot\beta +
    \partial_6\rho_s\nabla^2\dot\beta + \nonumber\\
\partial_1{\bf J}\cdot\nabla\alpha +
    \partial_2{\bf J}:\nabla^2\alpha +
    \partial_3{\bf J}\cdot:\nabla^3\alpha +
    \partial_4{\bf J}\cdot\nabla\beta +
    \partial_5{\bf J}:\nabla^2\beta +
    \partial_6{\bf J}\cdot:\nabla^3\beta -\nonumber\\
\Lambda_\alpha (\dot\alpha - g_\alpha)-\nonumber\\
\Lambda_{\nabla\alpha} (\nabla\dot\alpha -
  \partial_1g_\alpha\cdot\nabla\alpha -
  \partial_2g_\alpha:\nabla^2\alpha -
  \partial_3g_\alpha\cdot:\nabla^3\alpha -
  \partial_4g_\alpha\cdot\nabla\beta
  \partial_5g_\alpha:\nabla^2\beta -\nonumber\\
  \partial_6g_\alpha\cdot:\nabla^3\beta) -\nonumber\\
\Lambda_\beta (\dot\beta - g_\beta) -\nonumber\\
\Lambda_{\nabla\beta} (\nabla\dot\beta -
  \partial_1g_\beta\cdot\nabla\alpha -
  \partial_2g_\beta:\nabla^2\alpha -
  \partial_3g_\beta\cdot:\nabla^3\alpha -
  \partial_4g_\beta\cdot\nabla\beta -
  \partial_5g_\beta:\nabla^2\beta -\nonumber\\
  \partial_6g_\beta\cdot:\nabla^3\beta) \geq 0.
\end{gather}
Double and triple dots denote inner products of the corresponding
tensors (scalars are formed). One can group together the
coefficients by the derivatives of different orders
\begin{gather}
  \dot\alpha\left(\partial_1\rho_s - \Lambda_\alpha\right)+
  \dot\beta\left( \partial_4\rho_s - \Lambda_\beta \right) +
    \nonumber \\
  \nabla\dot\alpha \left(\partial_2\rho_s - \Lambda_{\nabla\alpha}\right) +
  \nabla\dot\beta \left(\partial_5\rho_s - \Lambda_{\nabla\beta}\right) +
    \nonumber \\
  \nabla^2\dot\alpha \partial_3\rho_s +
  \nabla^2\dot\beta \partial_6\rho_s +
    \nonumber\\
 \nabla^3\alpha \left(\partial_3{\bf J} +
    \Lambda_{\nabla\alpha}\partial_3 g_\alpha +
    \Lambda_{\nabla\beta}\partial_3 g_\beta \right) +
 \nabla^3\beta \left( \partial_6{\bf J} +
    \Lambda_{\nabla\alpha}\partial_6 g_\alpha +
    \Lambda_{\nabla\beta}\partial_6 g_\beta \right) +
    \nonumber \\
 \nabla^2\alpha \left(\partial_2{\bf J} +
    \Lambda_{\nabla\alpha}\partial_2 g_\alpha +
    \Lambda_{\nabla\beta}\partial_2 g_\beta \right) +
 \nabla^2\beta \left( \partial_5{\bf J} +
    \Lambda_{\nabla\alpha}\partial_5 g_\alpha +
    \Lambda_{\nabla\beta}\partial_5 g_\beta \right) +
    \nonumber \\
 \nabla\alpha \left(\partial_1{\bf J} +
    \Lambda_{\nabla\alpha}\partial_1 g_\alpha +
    \Lambda_{\nabla\beta}\partial_1 g_\beta \right) +
 \nabla\beta \left( \partial_4{\bf J} +
    \Lambda_{\nabla\alpha}\partial_4 g_\alpha +
    \Lambda_{\nabla\beta}\partial_4 g_\beta \right) +
    \nonumber \\
 \Lambda_\alpha g_\alpha + \Lambda_\beta g_\beta \geq 0.
\label{sl2}\end{gather} We find the Liu equations as the
coefficients of the derivatives that are not in the constitutive
space
 \bea
 \dot\alpha &:& \partial_1\rho_s - \Lambda_\alpha = 0, \label{liu1}\\
 \dot\beta &:& \partial_4\rho_s - \Lambda_\beta = 0, \label{liu2}\\
 \nabla\dot\alpha &:& \partial_2\rho_s - \Lambda_{\nabla\alpha} = {\bf 0},
    \label{liu3}\\
 \nabla\dot\beta &:& \partial_5\rho_s - \Lambda_{\nabla\beta} = {\bf 0},
    \label{liu4}\\
 \nabla^2\dot\alpha &:& \partial_3\rho_s = {\bf 0}, \label{liu5}\\
 \nabla^2\dot\beta &:& \partial_6\rho_s = {\bf 0}, \label{liu6}\\
 \nabla^3\alpha &:& \partial_3{\bf J} +
    \Lambda_{\nabla\alpha}\partial_3 g_\alpha +
    \Lambda_{\nabla\beta}\partial_3 g_\beta= {\bf 0}, \label{liu7}\\
 \nabla^3\beta &:& \partial_6{\bf J} +
    \Lambda_{\nabla\alpha}\partial_6 g_\alpha +
    \Lambda_{\nabla\beta}\partial_6 g_\beta= {\bf 0}. \label{liu8}
 \eea
The residual dissipation inequality follows from  Eq. \re{sl2} by
taking into account Eqs. \re{liu1}-\re{liu8}
 \begin{gather}
(\partial_1 {\bf J} + \Lambda_{\nabla\alpha}\partial_1 g_\alpha +
    \Lambda_{\nabla\beta}\partial_1 g_\beta)\cdot\nabla\alpha +
 (\partial_4 {\bf J} + \Lambda_{\nabla\alpha}\partial_1 g_\alpha +
    \Lambda_{\nabla\beta}\partial_4 g_\beta)\cdot\nabla\beta + \nonumber\\
 (\partial_2 {\bf J} + \Lambda_{\nabla\alpha}\partial_2 g_\alpha +
    \Lambda_{\nabla\beta}\partial_2 g_\beta):\nabla^2\alpha +
 (\partial_5 {\bf J} + \Lambda_{\nabla\alpha}\partial_5 g_\alpha +
    \Lambda_{\nabla\beta}\partial_5 g_\beta):\nabla^2\beta + \nonumber\\
 \partial_1 \rho_s g_\alpha + \partial_4 \rho_s g_\beta \geq 0.
 \label{resdis}\end{gather}
It is easy to see, that higher than first order derivatives of the
constraints do not give additional restrictions, due to the simple
structure of the evolution equations \re{d1}-\re{d2}.

The solution of the Liu equations \re{liu1}-\re{liu6} leads to the
entropy density, which is independent of the second gradients
$\rho_s(\alpha, \nabla \alpha, \nabla^2\alpha, \beta, \nabla\beta,
\nabla^2\beta) = \hat{\rho}_s (\alpha, \nabla \alpha, \beta,
\nabla\beta)$, and its partial derivatives are the Lagrange-Farkas
multipliers:
 $$
 \Lambda_\alpha = \partial_1 \rho_s =
    \frac{\partial \hat\rho_s}{\partial \alpha}, \quad
 \Lambda_{\nabla\alpha} = \partial_2 \rho_s =
    \frac{\partial  \hat\rho_s}{\partial \nabla\alpha}, \quad
 \Lambda_\beta = \partial_4 \rho_s =
    \frac{\partial  \hat\rho_s}{\partial \beta}, \quad
 \Lambda_{\nabla\beta} = \partial_5 \rho_s =
    \frac{\partial  \hat\rho_s}{\partial \nabla\beta}.
$$
Substituting the Lagrange-Farkas multipliers into Eqs. \re{liu7} and
\re{liu8}, one can solve them and get the entropy flux
\begin{gather}
 {\bf J}(\alpha, \nabla
\alpha, \nabla^2\alpha, \beta, \nabla\beta, \nabla^2\beta)=
    -\partial_2\rho_s g_\alpha - \partial_5\rho_s g_\beta +
    {\bf j}_0 = \nonumber\\
-\frac{\partial  \hat{\rho}_s}{\partial \nabla\alpha} g_\alpha -
 -\frac{\partial  \hat{\rho}_s}{\partial \nabla\beta} g_\beta
 + {\bf j}_0.
\label{ecd} \end{gather} Here the additional entropy flux term ${\bf
j}_0$ is a function only of the variables $(\alpha, \nabla \alpha,
\beta, \nabla\beta)$, as well as $\rho_s$. Substituting this
expression into the dissipation inequality,  we get finally:
 \begin{gather}
  \sigma_s = (\partial_1 \rho_s
    -\nabla\cdot \partial_2 \rho_s)g_\alpha
    +(\partial_4 \rho_s - \nabla\cdot \partial_5 \rho_s)g_\beta
    +\nabla\cdot {\bf j_0} =\nonumber\\
 \left(\frac{\partial  \hat{\rho}_s}{\partial \alpha}-
        \nabla\cdot\frac{\partial  \hat{\rho}_s}{\partial
            \nabla\alpha}\right)g_\alpha +
 \left(\frac{\partial  \hat{\rho}_s}{\partial \beta}-
        \nabla\cdot\frac{\partial  \hat{\rho}_s}{\partial
            \nabla\beta}\right)g_\beta  +\nabla\cdot {\bf j_0} \geq 0.
 \label{dif}\end{gather}
Let us remember, that $\hat{\rho}_s$ is independent of the second
gradients of the variables, therefore there are no higher
derivatives in Eq. \re{dif}. Let us assume now that ${\bf j}_0 =
{\bf 0}$. This is an assumption that we regularly consider in
classical irreversible thermodynamics excluding the appearance of
current multipliers and terms similar to that appeared in the
Guyer-Krumhansl equation \cite{Van01a2,Van03a}. We may recognize an
Onsagerian force-flux system, where the thermodynamic fluxes and
forces are
 \vskip 3mm
\begin{center}\begin{tabular}{cccc}
$\alpha$-force:  \hfill &
    $A = -\frac{\partial \rho_s}{\partial\alpha} +
        \nabla\cdot \frac{\partial \rho_s}{\partial \nabla\alpha}$  &
    $\qquad\alpha$-flux:  \hfill &
    ${g}_\alpha$ \hfill\\
$\beta$-force:  \hfill &
    $B = -\frac{\partial \rho_s}{\partial\beta} +
        \nabla\cdot \frac{\partial \rho_s}{\partial \nabla\beta}$ &
    $\qquad\beta$-flux: \hfill &
    ${g}_\beta$ \hfill
 \end{tabular}\end{center}\vskip 3mm
Here we introduced a shortened notation $A$ and $B$ for the
thermodynamic forces preserving the sign conventions of the previous
sections.

\subsection{Evolution equations}

Solution of the inequality (\ref{dif}) can be represented as
 \be
 \begin{pmatrix}g_\alpha \\ g_\beta \end{pmatrix} = -{\bf L}
 \begin{pmatrix}A \\B \end{pmatrix} =
    \begin{pmatrix} l_1 & l_{12} \\ l_{21}  & l_2 \end{pmatrix}
    \begin{pmatrix} -A \\ -B \end{pmatrix}.
 \ee
This corresponds to the inequality (\ref{dif}) if the Onsagerian
coefficients $l_1,l_2,l_{12}, l_{21}$ are functions of the
thermodynamic forces $A$ and $B$ and the state variables $(\alpha,
\nabla\alpha, \beta,$ $\nabla\beta)$. Signs of the coefficients
should be further restricted by the requirement of nonnegative
entropy production.

In what follows, we split the conductivity matrix into a symmetric
and an antisymmetric part, introducing $k$ and $l$ instead of
$l_{12}$ and $l_{21}$ by $l_{12} = l-k$ and $l_{12} = l+k$:
 \bea
 \dot\alpha &=& g_\alpha =
    k B - l_1 A - l B \label{o1} \\
 \dot\beta &=& g_\beta =
    -k A  - l A - l_2 B
    \label{o2}.
 \eea
The nonnegativity of the entropy production \re{dif} results in the
positive semidefiniteness of the symmetric part of the conductivity
matrix {\bf L}. Therefore, the following inequalities are valid for
the coefficients \be l_1\geq0,\quad  l_2\geq0,  \quad l_1 l_2 -
l^2\geq0. \ee

\subsection{Remark on dissipation potentials}

We may introduce dissipation potentials for the dissipative part of
the equations, if the condition of their existence is satisfied.
However, in the considered case there is no need of this assumption,
because here it is clear what belongs to the dissipative part and
what belongs to the nondissipative part of the the evolution
equations. The terms with the symmetric conductivity contribute to
the entropy production and the terms from the skew symmetric part do
not. On the other hand,  there is no need of potential construction,
as we are not looking for a variational formulation. Moreover,  the
symmetry relations are not sufficient for the existence of
dissipation potentials in general, as we have emphasized previously.
In the case of constant coefficients (strict linearity), the
dissipation potentials always exist for the dissipative (symmetric)
part.

\subsection{Remark on the reciprocity relations}

The reciprocity relations are the main results of the great idea of
Lars Onsager connecting fluctuation theory to macroscopic
thermodynamics \cite{Ons31a1,Ons31a2,OnsMac53a,MacOns53a,Cas45a}. As
it was written by Onsager himself on the validity of his result:
"The restriction was stated: on a kinetic model, the thermodynamic
variables must be algebraic sums of (a large number of) molecular
variables, and must be {\em even} functions of those molecular
variables which are odd functions of time (like molecular
velocities)" \cite{MacOns53a}. The Casimir reciprocity relations are
based on microscopic fluctuations, too \cite{Cas45a}. We do not have
such a microscopic background for most of internal variables. E. g.,
in the case of damage we think that the internal variables are
reflecting a structural disorder on a mesoscopic scale. The relation
between thermodynamic variables and the microscopic structure is
hopelessly complicated. On the other hand, the Onsagerian
reciprocity is based on time reversal properties of corresponding
physical quantities either at the macro or at the micro level.
Looking for the form of evolution equations without a microscopic
model, we do not have any information on the time reversal
properties of our physical quantities  neither at the micro- nor at
the macroscopic level. Therefore, we can conclude that lacking the
conditions of the Onsagerian or Casimirian reciprocity gives no
reasons to assume their validity in the internal variable theory.

Let us observe the correspondence of evolution equations for
internal variables with the reciprocity relations by means of a few
simple examples.

\subsection{Example 1: internal variables of state}

Let us consider materials with zero antisymmetric part of the
conductivity matrix ${\bf L}$  ($l = 0$, $k=0$). It is clear that
the Onsagerian reciprocity relations are satisfied, and we return to
the classical situation with fully uncoupled internal variables:
 \bea
 \dot\alpha &=& g_\alpha =
    -l_1 A , \\
 \dot\beta &=& g_\beta =
    - l_2 B
   .
 \eea
In this case the evolution equations for dual internal variables
$\alpha$ and $\beta$ are the same as in the case of single internal
variable of state (Eq. \re{GLev}).

\subsection{Example 2: internal degrees of freedom}
\label{idf}

We now assume that all conductivity coefficients are constant, and
their values are $l_1 = l = 0$, $k=1$. This means that
$l_{12}=-l_{21}$, i.e., the Casimirian reciprocity relations are
satisfied.

For simplicity, we decompose the entropy density into two parts,
which are dependent on different internal variables
 \be
{\rho_s}(\alpha,\nabla\alpha, \beta,\nabla\beta) = -K(\beta) -
W(\alpha, \nabla\alpha).
 \label{p}\ee
The negative signs are introduced taking into account the concavity
of the entropy. Then the thermodynamic forces are represented as
$$
 A = \frac{\partial W}{\partial\alpha} -
    \nabla\cdot \frac{\partial W}{\partial \nabla\alpha}, \qquad
 B = \frac{d K}{d\beta} = K'(\beta),
 $$
and Eqs. \re{o1}-\re{o2} are simplified to
 \bea
 \dot\alpha &=& g_\alpha = B = K'(\beta)  \label{so1} \\
 \dot\beta &=& g_\beta = -A + l_2 B =
    - \frac{\partial W}{\partial\alpha} +
    \nabla\cdot \frac{\partial W}{\partial \nabla\alpha}- l_2K'(\beta)
    \label{so2}.
 \eea
One may recognize that the obtained system of equation corresponds
exactly to what was introduced in the case of dynamic degrees of
freedom after transforming the equation of motion into a Hamiltonian
form \re{H1} and \re{H2}.

The transformation into a Lagrangian form is trivial if $K$ has a
quadratic form $K(\beta) = \frac{\beta^2}{2 m}$, where $m$ is a
constant. Then we get exactly Eqs. \re{sh1}-\re{sh2} and the whole
system corresponds to Eq. (5.14) in \cite{MauMus94a1}  with the
Lagrangian
 $$
  \mathcal{L}(\dot\alpha, \alpha, \nabla\alpha) =
    m\frac{\dot\alpha^2}{2} - W(\alpha,\nabla\alpha),
  \quad \text{and} \quad
  D(\alpha, \dot{\alpha}) =  \frac{ml_2}{2}\dot\alpha^2.
$$
Moreover, the entropy flux density \re{ecd} in the case of our
special conditions can be written as
 \be
 {\bf J}=
    -\frac{\partial  \rho_s}{\partial \nabla\alpha} K'(\beta)
    + {\bf j}_0.
 \ee
and one can infer that natural boundary conditions of the
variational principle \re{bou} correspond to the condition of
vanishing entropy flow at the boundary, i.e.,  ${\bf j}_0={\bf 0}$.

Therefore, the variational structure of internal degrees of freedom
is recovered in the pure thermodynamic framework. The thermodynamic
structure resulted in several sign restrictions of the coefficients,
and the form of the entropy flux is also recovered. The natural
boundary conditions correspond to a vanishing extra entropy flux.

\subsection{Example 3: diffusive internal variables}

Now we give an additional example to see clearly the reduction of
evolution equations of internal degrees of freedom to evolution
equations for internal variables and the extension of the later to
the previous one.

We keep the values of conductivity coefficients (i.e., $l_1 = l =
0$, $k=1$), but assume that both $K$ and $W$ are quadratic functions
$$
K(\beta)=\frac{b}{2}\beta^{2}, \quad W(\alpha,
\nabla\alpha)=\frac{a}{2}(\nabla\alpha)^{2},
$$
where $a$ and $b$ are positive constants according to the concavity
requirement.

In this case, the evolution equations (\ref{so1}), (\ref{so2})
reduce to
 \bea
 \dot\alpha &=& g_\alpha = B = K'(\beta)=b\beta,  \label{so3} \\
 \dot\beta &=& g_\beta =
    - \frac{\partial W}{\partial\alpha} +
    \nabla\cdot \frac{\partial W}{\partial \nabla\alpha}-
    l_2K'(\beta)=a\Delta \alpha -l_2b\beta
    \label{so4}.
\eea Substituting $\beta$ from Eq. (\ref{so3}) into Eq. (\ref{so4}),
we have
 \be
  \frac{1}{ab}\ddot\alpha + \frac{l_2}{a}\dot\alpha=\Delta \alpha
    \label{so5}.
  \ee
which is a Cattaneo-Vernotte type hyperbolic equation (telegraph
equation) for the internal variable $\alpha$. This can be considered
as an extension of a diffusion equation by an inertial term or as
and extension of a damped Newtonian equations (without forces) by a
diffusion term.

\section{Discussion}

In the framework of the thermodynamic theory with dual weakly
nonlocal internal variables presented in the paper, we are able to
recover the evolution equations for internal degrees of freedom.

We have seen that the form of evolution equations depends on the
mutual interrelations between the two internal variables. In the
special case of internal degrees of freedom, the evolution of one
variable is driven by the second one, and vice versa. This can be
viewed as a duality between the two internal variables. In the case
of pure internal variables of state, this duality is replaced by
self-driven evolution for each internal variable. The general case
includes all intermediate situations.

It is generally accepted that internal variables are "measurable but
not controllable" (see e.g. \cite{Kes93a1}). Controllability can be
achieved by boundary conditions or fields directly acting on the
physical quantities. We have seen how natural boundary conditions
arise considering nonlocality of the interactions through weakly
nonlocal constitutive state spaces.

As we wanted to focus on generic inertial effects, our treatment is
simplified from several points of view. Vectorial and tensorial
internal variables were not considered and the couplings to
traditional continuum fields result in degeneracies and more
complicated situations than in our simple examples.

A number of systems with these degeneracies are called {\em
gyroscopic}. One can distinguish a least two types of them. In
thermodynamics this is a customary nomination for skew symmetric
Casimir type couplings of thermodynamic interactions, because in
this case the entropy production is zero. For example a single
second order tensor internal variable $\boldsymbol \xi$ can couple
to mechanical interactions for isotropic materials, too. In this
case the relevant part of the entropy production from Eq. \re{bale}
can be written as:
 \be
 T \sigma_s = ({\bf t}-{\bf t}^{rev}) : \nabla{\bf v} -
    {\boldsymbol \Xi}:\dot{\boldsymbol  \xi} \geq 0.
\ee Here $\boldsymbol \Xi= -\rho \frac{\partial s}{\partial \xi}$ is
the related affinity. If $\boldsymbol \xi$ is a variable that
changes sign under time reversal then there is a pure skew symmetric
coupling between the two variables and the entropy production is
zero. This is a valid and important example in rheology (see
\cite{Ver97b} for second order traceless tensors as internal
variables). The above example is sometimes called as gyroscopic
(e.g. \cite{MauDro83a}), because the corresponding thermodynamic
forces and fluxes are orthogonal.

On the other hand the gyroscopic systems in mechanics are typically
characterized by a Lagrangian with the following scalar product
term:
 \ben
  L({\bf q, \dot q}) = {\bf a}({\bf q}) \cdot {\bf q},
 \een
\noindent where ${\bf a}$ is a vector field, that depends on the
generalized coordinates ${\bf q}$ of the mechanical system and has
the same dimension as coordinate space. The dot denotes the scalar
product. The corresponding Euler-Lagrange equation follows:
 \ben
  D\wedge {\bf a}\cdot {\bf q} = {\bf 0}.
 \een
Here the derivative  $D\wedge {\bf a} = D{\bf a} - (D{\bf a})^T$.
This system is degenerate, because the corresponding Legendre
transformation is not invertible, the related Hamiltonian is
identically zero.

These two examples show well that skew symmetric couplings are not
always related directly to inertial effects and indicate two
directions, one into mechanics and one into thermodynamics, where
our method can be generalized. Let us mention here the related
pioneering works of Verh\'as, where skew symmetric conductivity
equations appear in different inspiring contexts
\cite{Ver89a,Ver01p}.

Finally, let us mention that the idea of constructing a unified
theoretical frame for reversible and irreversible dynamics has a
long tradition. The corresponding research was not restricted to the
case of internal variables and was looking for a classical
Hamiltonian or a generalized variational principle that would be
valid for both dissipative and nondissipative evolution equations
(see, e.g., \cite{Gya70b,GlaPri71b,MarGam91a,VanNyi99a} and the
references therein).

\section*{Acknowledgements}

This research was supported by OTKA T048489, by the EU-I3HP project
and the Bolyai scholarship of Peter V\'an. The support from the
Estonian Science Foundation (A.B. and J.E) is acknowledged. All
authors are grateful to Marie Curie Transfer of Knowledge project
CENS-CMA.

\bibliographystyle{unsrt}

\end{document}